# Electrical Detection of Spin Transport in Lateral Ferromagnet-Semiconductor Devices

X. Lou,[1] C. Adelmann,[2] S.A. Crooker,[3] E. S. Garlid,[1] J. Zhang,[1] S.M. Reddy,[2] S.D. Flexner,[2] C.J. Palmstrøm,[2] and P.A. Crowell[1]

[1] School of Physics and Astronomy, University of Minnesota, Minneapolis, MN 55455.

[2] Department of Chemical Engineering and Materials Science, University of Minnesota, Minneapolis, MN 55455.

[3] National High Magnetic Field Laboratory, Los Alamos, NM 87545.

**A longstanding goal of research in semiconductor spintronics is the ability to inject, modulate, and detect electron spin in a single device[1-4]. A simple prototype consists of a lateral semiconductor channel with two ferromagnetic contacts, one of which serves as a source of spin-polarized electrons and the other as a detector. Based on work in analogous metallic systems [5-8], two important criteria have emerged for demonstrating electrical detection of spin transport. The first is the measurement of a non-equilibrium spin population using a "non-local" ferromagnetic detector through which no charge current flows[5,7]. The potential at the detection electrode should be sensitive to the relative magnetizations of the detector and the source electrodes, a property referred to as the spin-valve effect. A second and more rigorous test is the existence of a Hanle effect, which is the modulation and suppression of the spin valve signal due to precession and dephasing in a transverse magnetic field [5,8]. Here we report on the observation of both the spin valve and Hanle effects in lateral devices consisting of epitaxial Fe Schottky tunnel barrier contacts on an *n*-doped GaAs channel. The dependence on transverse magnetic field, temperature, and contact separation are in good agreement with a model incorporating spin drift and diffusion. Spin transport is detected for both directions of current flow through the source electrode. The sign of the electrical detection signal is found to vary with the injection current and is correlated with the spin polarization in the GaAs channel determined by optical measurements. These**

**results therefore demonstrate a fully electrical scheme for spin injection, transport, and detection in a lateral semiconductor device.**

A schematic of the spin transport devices used for these experiments is shown in Fig. 1(a). The ferromagnet-semiconductor (FS) devices are fabricated from epitaxial Fe/GaAs (100) heterostructures [9,10]. The semiconductor channel is a lightly n-doped GaAs epilayer ($n = 2 – 4 \times 10^{16}$ cm$^{-3}$, 2500 nm thick), and a Schottky tunnel barrier is formed at the interface between the Fe (5 nm thick) and the GaAs by growing a $n^+$-doped (~ $5 \times 10^{18}$ cm$^{-3}$) GaAs transition layer[11]. The presence of the Schottky tunnel barriers allows for efficient electrical spin injection and detection [12,13]. Standard photolithography and etching techniques are used to define the channel and to pattern the five Fe electrodes, which have nominal dimensions of 10 μm × 50 μm. The three central contacts have a center-to-center spacing of 12 μm, and the two end contacts are 160 μm from the three central contacts. The magnetic easy axes of the Fe contacts are along the GaAs [011] direction, which is parallel to the long axes of the contacts. Measurements on devices fabricated from three different heterostructures, denoted A, B, and C, will be discussed in this paper. The channel dopings determined by Hall measurements at 10 K are $n = 3.5 \times 10^{16}$ cm$^{-3}$, $2.0 \times 10^{16}$ cm$^{-3}$, and $2.5 \times 10^{16}$ cm$^{-3}$ for A, B, and C respectively.

The interpretation of electrical spin transport measurements can be complicated by magnetoresistance in the electrodes, local Hall effects, and other extrinsic contributions to the signal[14]. A "non-local" measurement[6] minimizes these background effects by placing a spin detection electrode outside the path of the charge current. The geometry is shown in Fig. 1(a). Spin-polarized electrons are injected into the GaAs channel at contact $c$ and flow towards contact $a$, while the voltage $V_{de}$ is measured between contacts $d$ and $e$. Although electrons flow from $c$ to $a$, the non-equilibrium spin polarization in GaAs, represented by the purple arrows in Fig. 1(a), can diffuse in either direction from the source. The spin polarization $P_{GaAs}$ results in an electrochemical potential difference $\Delta\mu$ for the two spin states in the channel, leading to a change in $V_{de}$ when the magnetization of $d$ is switched from antiparallel to $c$ to the parallel configuration. $P_{GaAs}$ at contact $e$, which is 160 μm away, is always zero, and so the magnetization of $e$ does not affect the measurement. The spin-valve measurement is carried out by sweeping the magnetic field along the magnetic easy axis [ $\hat{y}$ in Fig. 1 (a)], and looking for a change in voltage over the narrow field range in which contacts $c$ and $d$ are antiparallel.

Measurements of $V_{de}$ are shown in the top panel of Fig. 1(b) for a field sweep at a current $I_{ac}$ = 1.0 mA (electrons flowing from $c$ to $a$) at $T$ = 50 K. The raw data shown in the top panel include an offset $V_0$ = -30.227 mV (much larger than the spin-dependent effects) resulting from spreading of the charge current in the GaAs channel as well as background contributions that are linear and quadratic in magnetic field. The background is fitted and subtracted from the raw data, yielding the curves shown in the bottom panel of Fig. 1(b). The two square jumps with a magnitude of 16.8 ± 0.2 µV occur over the field range in which the magnetizations of contacts $c$ and $d$ are antiparallel.

The data of Fig. 1(b) indicate the existence of a lateral spin valve effect. The interpretation of spin-valve measurements on FS devices in the two-terminal geometry[15] has been subject to challenge[14], and previous non-local measurements[16,17] have not observed clear switching signatures such as those in Fig. 1(b). Most importantly, however, previous measurements on FS devices have not demonstrated precession of the spin between the source and detection electrodes. The simplest manifestation of this property is the Hanle effect[5,8], in which the magnetic field-dependence of the non-local voltage is due to precession and dephasing of the spins in the semiconductor. The precession is induced by applying a small *transverse* magnetic field that does not change the magnetizations of the electrodes. To test for a Hanle effect in our devices, the magnetizations of contacts $c$ and $d$ were set in the parallel state. The magnetic field perpendicular to the plane ( $B_z$ in Fig. 1) was then swept, resulting in the black data points shown in the top panel of Fig. 1(c). The offset $V_0$ is the same as for Fig. 1(b). The corresponding data after subtraction of the background (dashed blue line) are shown in the lower panel of Fig. 1(c). This procedure was repeated after setting $c$ and $d$ into the antiparallel state, yielding the red points in Fig. 1(c). These data show a peak at $B_z$ = 0 rather than the minimum observed in the parallel state. As can be seen in the raw data, the two Hanle curves merge at large $B_z$, since in this limit the spins in the GaAs channel are completely dephased. The difference in the two Hanle signals at $B$ = 0 is 18.0 ± 0.1 µV, differing from the jump in the spin valve data by 1.2 µV.

The data of Fig. 1 therefore demonstrate a completely electrical scheme for spin injection, transport, and detection in a lateral semiconductor device. We now consider a quantitative interpretation of the Hanle data using a drift-diffusion model developed previously[5,8]. Consider spins, oriented along $\hat{y}$ in the coordinate system of Fig. 1(a), which are electrically injected at a point $x_1$ and diffuse towards a detector located at $x_2$. While they are diffusing, the electron spins relax at a rate $\tau_s^{-1}$ and precess about the applied field $\vec{B} = B\hat{z}$ at the Larmor frequency $\Omega = g\mu_B B/\hbar$, where $g$ is the electron g-factor, $\mu_B$ is the Bohr magneton, and $\hbar$ is Planck's constant. The y-component steady-state spin polarization at the detector is[10]

$$S_y(x_1, x_2, B) = S_0 \int_0^\infty \frac{1}{\sqrt{4\pi Dt}} e^{-(x_2-x_1-v_d t)^2/4Dt} \cos(g\mu_B Bt/\hbar) e^{-t/\tau_s} dt, \qquad (1)$$

where $S_0$ is the spin injection rate and $D$ is the electron diffusion constant. The drift velocity $v_d$ is zero for purely diffusive transport. $S_y(x_1, x_2, B)$ is then integrated over the widths of the source and detector to obtain $S_y(B)$, which is proportional to the spin-dependent non-local voltage. The solid curves in Fig. 2, which shows Hanle data (after background subtraction) for sample B at temperatures from 10 to 70 K, are fits to Eq. (1), with $D$ determined from the measured carrier mobility and charge density. The g-factor is fixed at $g = -0.44$, and $S_0$ and $\tau_s$ are free parameters. The increase in width and decrease in amplitude of the Hanle curves with increasing $T$ are due primarily to the decrease in $\tau_s$ from 24 nsec at 10 K to 4 nsec at 70 K.

The sample design shown in Fig. 1(a) allows for different permutations of the injection and detection contacts. The first two possibilities, in which the electrons flow from contact b or c to contact a while the voltage $V_{de}$ is measured, correspond to spin transport by diffusion alone. Hanle curves for these two cases measured on sample A at 50 K and a current of 1.0 mA are shown (after background subtraction) in Fig. 3(a). The two panels show data for source-detector separations $\Delta x = +12$ μm and $+24$ μm, and the black and red points are taken for the parallel and antiparallel configurations respectively. In addition, it is possible to place the detection contact in the current path, so that spins are injected at contact d and detected at either c or b. In this case, which we refer to as the crossed configuration, the drift velocity $v_d$ in Eq. 1 is non-zero. Data obtained for the two crossed measurements ($\Delta x$ = -12 and -24 µm) are shown in Fig. 3(b).   Given the four possible contact separations and two magnetization configurations, eight different Hanle curves were obtained for a single bias current.   Fits of the data to Eq. 1 are shown as the solid curves in Fig. 3(a) and 3(b). We model the data with the same spin injection rate $S_0$ for all eight cases.  The diffusion constant $D$, drift velocity $v_d$, and spin lifetime $\tau_s$ were determined independently from magneto-transport and optical measurements. Given the large number of constraints, the agreement between the model [solid curves in Figs. 3(a) and (b)] and the data is very good, with significant deviations only for the case $\Delta x$ = -24 µm.   The decrease in the widths of the Hanle curves with increasing $\Delta x$ and the more prominent minima at non-zero field (due to precession) reflect the fact that the average time for spins to reach the detector is longer.    The signal is largest for $\Delta x$ = -12 µm, for which the average time for electrons to reach the detector is shortest. The peak-to-peak amplitude $V_{\uparrow\downarrow} - V_{\uparrow\uparrow}$ of the Hanle signal is shown as a function of $\Delta x$ in Fig. 3(c), along with curves generated from the drift-diffusion model.   The magnitudes obtained from Hanle curves (triangles) and spin-valve measurements (circles) agree within 1.5 µV in all cases.

We now consider the magnitude of the electrical spin detection signal.  Following the interpretation for metallic systems[5], we consider a ferromagnetic voltage probe with spin polarization (at the Fermi level) $P_{Fe}$, which is in contact with spin-polarized electrons in GaAs.  The voltage for the two different magnetization states of the probe is measured with respect to unpolarized GaAs far from the contact.  The $n$-GaAs is modeled as a Pauli metal with a carrier density $n = 3 \times 10^{16} \text{cm}^{-3}$ and effective mass $0.07 m_e$. The spin polarization in the semiconductor is $P_{GaAs} = (n_\uparrow - n_\downarrow)/(n_\uparrow + n_\downarrow)$, where $n_\uparrow$ and $n_\downarrow$ are the total densities in each of the two spin bands.  The corresponding electrochemical potential difference in the limit $\Delta\mu << E_f$ is $\Delta\mu = (n_\uparrow - n_\downarrow)/N(E_f)$, where $N(E_f)$ is the density of states at the Fermi energy in GaAs.   The conditions for electrochemical equilibrium lead to a voltage difference[5]

$$V_{\uparrow\downarrow} - V_{\uparrow\uparrow} = \frac{\eta P_{Fe}}{e}\Delta\mu = \frac{2\eta P_{Fe} P_{GaAs} E_f}{3e}, \qquad (2)$$

where $\eta$ is the spin transmission efficiency of the interface and $e$ is the electronic charge. We assume $P_{Fe} = 0.42$ [18] and $\eta \sim 0.5$ [19]. From Eq. 2, a signal $V_{\uparrow\downarrow} - V_{\uparrow\uparrow} = 15$ μV at the detector [$\Delta x = +12$ μm in Fig. 3(c)] corresponds to $P_{GaAs} = 0.02$. Given the measured spin diffusion length, $l_D = \sqrt{D\tau_s} = 6$ μm at 50 K, this result implies $P_{GaAs} = 0.16$ at the source ($\Delta x = 0$), which is close to value $\eta P_{Fe} \sim 0.2$ assumed for the injected polarization.

The data in Figs. 1 – 3 were obtained with electrons flowing from the Fe source contact into GaAs (Schottky barrier reverse-biased), but it is also possible to generate a spin accumulation when electrons flow from GaAs into Fe (forward bias)[9,10,20,21]. Spin-valve measurements for sample C obtained for both directions of bias current are shown in Fig. 4(a). The top two curves were obtained at currents of +0.02 and +0.01 mA, corresponding to injection of electrons from Fe into GaAs. In this case, $V_{\uparrow\downarrow} - V_{\uparrow\uparrow}$ is positive. At small negative currents, $V_{\uparrow\downarrow} - V_{\uparrow\uparrow}$ is negative, as expected in the linear response regime. At larger negative currents, however, $V_{\uparrow\downarrow} - V_{\uparrow\uparrow}$ goes through a minimum, switches sign at -60 μamps and is then positive at all larger negative bias currents. (The cusps at zero field in these data track $V_{\uparrow\downarrow} - V_{\uparrow\uparrow}$ but are strongly dependent on the sweep rate and temperature. They are probably due to hyperfine effects[22].) The results of these measurements in the low-bias regime are summarized as filled squares in Fig. 4(b), which shows $V_{\uparrow\downarrow} - V_{\uparrow\uparrow}$ as a function of the interfacial voltage $V_{int}$, where $V_{int}$ is the voltage drop measured between the channel and the source contact (see Supplementary Figure 1 for a typical *I-V* curve).

The bias dependence shown in Figs. 4(a) and (b) raises the question of how the non-local voltage is related to the electron spin polarization $P_{GaAs}$ in the semiconductor channel at different bias conditions. To address this issue, we have measured $P_{GaAs}$ directly using the magneto-optical Kerr effect, exploiting the fact that the Kerr rotation $\theta_K$ is proportional to $P_{GaAs}$. Each device was placed in a low-temperature magneto-optical Kerr microscope, and $\theta_K$ at a position in the channel ~ 8 μm from contact *a* was measured as a function of bias using the technique of Ref. 9. The Kerr rotation data for $P_{GaAs}$ in sample C are shown in Fig. 4(b) (red circles) overlayed on the data for $V_{\uparrow\downarrow} - V_{\uparrow\uparrow}$.

At low bias, for which drift effects can be neglected, $P_{GaAs}$ tracks $V_{\uparrow\downarrow} - V_{\uparrow\uparrow}$, with a minimum at -40 mV and a sign reversal at -100 mV. The sign of $P_{GaAs}$ is the same at large positive and negative $V_{int}$ and corresponds to *majority* spin polarization in Fe (spin polarization opposite to magnetization) [9].

The bias-dependences of the non-local voltage $V_{\uparrow\downarrow} - V_{\uparrow\uparrow}$ and the polarization $P_{GaAs}$ have been measured for all of the devices discussed in this paper. The measurements on sample A (Supplementary Figure 2) are similar to those on sample C, but $V_{\uparrow\downarrow} - V_{\uparrow\uparrow}$ for sample B is *opposite* in sign to the other samples at large bias (for both current directions). As shown in Fig. 4(c), $V_{\uparrow\downarrow} - V_{\uparrow\uparrow}$ approaches zero and passes through a maximum at small positive $V_{int}$. The opposite sign of $V_{\uparrow\downarrow} - V_{\uparrow\uparrow}$ in sample B is observed in spite of the fact that, as for the other samples at high bias, $P_{GaAs}$ is positive (majority spin polarization) outside of the range $0 < V_{int} < 10$ mV. In this case, $P_{GaAs}$ overlaps almost perfectly with the non-local signal after the sign of $V_{\uparrow\downarrow} - V_{\uparrow\uparrow}$ is inverted [open squares in Fig. 4(c)]. This observation is directly related to the most significant difference in $P_{GaAs}$ among the samples, which is the location of the minimum near zero bias and therefore the slope $\partial P_{GaAs}/\partial V_{int}$ at $V_{int} = 0$. This derivative determines (by reciprocity) the voltage $V_{\uparrow\downarrow} - V_{\uparrow\uparrow}$ that is measured at the unbiased detection contact due to a spin polarization $P_{GaAs}$ [5]. As can be seen from the Kerr rotation data in Figs. 4(b) and (c), $(\partial P_{GaAs}/\partial V_{int})_{V_{int}=0}$ is positive for sample C but is negative for sample B. The sign of the non-local voltage for a given $P_{GaAs}$ should therefore be opposite for sample B, as is observed.

The optical measurements of $P_{GaAs}$ are therefore consistent with the non-local voltage measurements, although we do not have an explanation for why the position of the minimum in $P_{GaAs}$ vs. $V_{int}$ changes from sample to sample. A simple argument based on the spin-polarized density of states of Fe predicts that a minimum should occur at negative $V_{int}$ (forward bias), although this is not necessarily consistent with tunneling measurements on vertical FS structures[23]. It is likely that the actual tunneling matrix elements are sensitive to details of the Schottky barrier profile or the interfacial band structure of Fe [24-26]. For example, if the primary contributions to the tunneling current for the two spin-bands occur at different wave-vectors [24,27,28], then the matrix elements for different spins can vary depending on the barrier height or width. Other processes, such as tunneling from bound states near the interface, may also need to be considered [29]. As demonstrated here, the study of these transport processes is advanced significantly by the ability to combine spin injection, modulation, and detection in a single electrical ferromagnet-semiconductor device.

**Methods**

All three samples were grown on (100) semi-insulating GaAs substrates, with epitaxial layers consisting of (in growth order) a 300 nm undoped GaAs buffer layer, a 2500 nm Si-doped $n$-GaAs layer (the channel), a 15 nm $n \rightarrow n^+$ GaAs transition layer, 15 nm $n^+$ GaAs, 5 nm Fe, and a 3 nm Al capping layer. The $n^+$ layer, for which the nominal doping is $5 \times 10^{18}$ cm$^{-3}$ in all three samples, forms a narrow Schottky barrier.

Wet-etching or ion milling was used to define the 10 μm × 50 μm Fe contacts. A 374 μm × 70 μm channel was defined by wet etching down to the substrate. The $n^+$ and $n \rightarrow n^+$ transition layers were then removed by wet etching so that the current was confined to the $n$-doped GaAs channel. A 200 nm SiN isolation film layer was then deposited at 100 °C. Finally, Ti/Au vias and bonding-pads were fabricated by electron beam evaporation and liftoff. The coercivities of the contacts were typically 150 – 300 Oe. The perpendicular anisotropy field for Fe is 2 T, and so the small fields $B_z$ applied in the Hanle measurements have a negligible effect on the magnetization of the Fe contacts.

The non-local measurements were carried out using a current source and nanovoltmeter. The carrier concentration and mobility were measured on companion Hall structures. Spin lifetimes used in the modeling were determined from Hanle curves obtained under optical pumping[30]. Optical measurements of the spin polarization were carried out using magneto-optical Kerr microscopy as described in Ref. 9. The absolute spin orientation of electrically-injected electrons in GaAs was found by comparing the sign of the measured Kerr rotation with the sign of the Kerr rotation induced by optically-injected electrons (created using circularly polarized light at 1.58 eV) with a known spin direction.

**Figure Captions:**

**Figure 1**: **Schematic of the experiment and representative non-local spin-valve and Hanle effects. (a)** A schematic diagram of the non-local experiment (not to scale). The five 10 × 50 μm Fe contacts have magnetic easy axes along $\hat{y}$, which is the GaAs [011] direction. The large arrows indicate the magnetizations of the source and detector. The two different contact separations are $l_1$ = 160 μm and $l_2$ = 12 μm. Electrons are injected along the path shown in red. The injected spins (purple) diffuse in either direction from contact $c$. The non-local voltage is detected at contact $d$. Other choices of source and detector among contacts $b$, $c$, and $d$ are also possible. **(b)** Non-local voltage $V_{de}$ vs. in-plane magnetic field $B_y$ (swept in both directions) for sample A at a current $I_{ac}$ = 1.0 mA at $T$ = 50 K. Raw data are shown in the upper panel (with an offset $V_0$ = -30.227 mV subtracted). The background (dashed blue curve underneath the data) is fitted by a 2$^{nd}$ order polynomial. The lower panel shows the data with this background subtracted. **(c)** Non-local voltage $V_{de}$ vs. perpendicular magnetic field $B_z$ for the same contacts and bias conditions (and the same offset $V_0$) as in (b). Data in the lower panel have the background (dashed blue curve in upper panel) subtracted. The data shown in black are obtained with the magnetizations of $c$ and $d$ parallel, and the data shown in red are obtained in the antiparallel configuration.

**Figure 2: Hanle curves at different temperatures**. Hanle curves for sample B obtained from $V_{ce}$ ($\Delta V$ is $V_{ce}$ after background subtraction) for a current $I_{ab}$ = 0.6 mA at several different temperatures. The curves are offset for clarity. The solid curves are fits to the model described in the text.

**Figure 3: Dependence of the non-local signal on contact separation. (a)** Hanle curves (background subtracted) for sample A at $T$ = 50 K and $I$ = 1.0 mA. The upper panel shows data obtained for parallel (black) and antiparallel (red) magnetizations for a contact pair with a separation $\Delta x$ = +12 μm (source is $c$ and detector is $d$). The lower panel shows data for $\Delta x$ = +24 μm (source is $b$ and detector is $d$). **(b)** The corresponding Hanle curves for $\Delta x$ = -12 μm (upper panel) and -24 μm (lower panel). For these two

"crossed" contact configurations, the detection electrode is in the current path, as shown in the sketch below the data. The solid curves in (a) and (b) are fits to the model described in the text. **(c)** The voltage difference $V_{\uparrow\downarrow} - V_{\uparrow\uparrow}$ between the antiparallel and parallel configurations vs. contact separation for sample A at 50 K. Open circles show results obtained from the spin-valve geometry and the triangles show data obtained in the Hanle geometry. Solid curves are obtained from Eq. 1.

**Figure 4**: **Bias dependence of the non-local signal and the spin polarization.**
**(a)** Non-local voltage $V_{de}$ ( $\Delta V$ is $V_{de}$ after background subtraction, with an offset for clarity) measured in sample C for different bias currents $I_{ac}$ in the in-plane geometry. Reverse bias data ($I_{ac} > 0$, for which electrons flow from $c$ to $a$) are shown in black; forward bias data ($I_{ac} < 0$) are shown in blue. **(b)** $V_{\uparrow\downarrow} - V_{\uparrow\uparrow}$ (black squares, left axis) near zero bias is shown vs. the interfacial voltage $V_{int}$ at the source contact. The Kerr rotation $\theta_K$ (red circles, right axis) vs. $V_{int}$ is measured on the same device. $\theta_K$ is proportional to the electron spin polarization $P_{GaAs}$ in the semiconductor. **(c)** $V_{\uparrow\downarrow} - V_{\uparrow\uparrow}$ (solid black squares, left axis) and $\theta_K$ (red circles, right axis) for sample B. The open black squares show the non-local voltage with the sign inverted (i.e. $V_{\uparrow\uparrow} - V_{\uparrow\downarrow}$). The region near zero is magnified in the inset (solid squares omitted).

**Acknowledgements** We thank B. D. Schultz and K. Raach for assistance, and E. Dan Dahlberg for useful discussions. This work was supported by the Office of Naval Research, the National Science Foundation MRSEC and NNIN Programs, and the Los Alamos LDRD program.

**Competing interests statement** The authors declare that they have no competing financial interests.

**Correspondence** and requests for materials should be addressed to P.A.C. (crowell@physics.umn.edu).

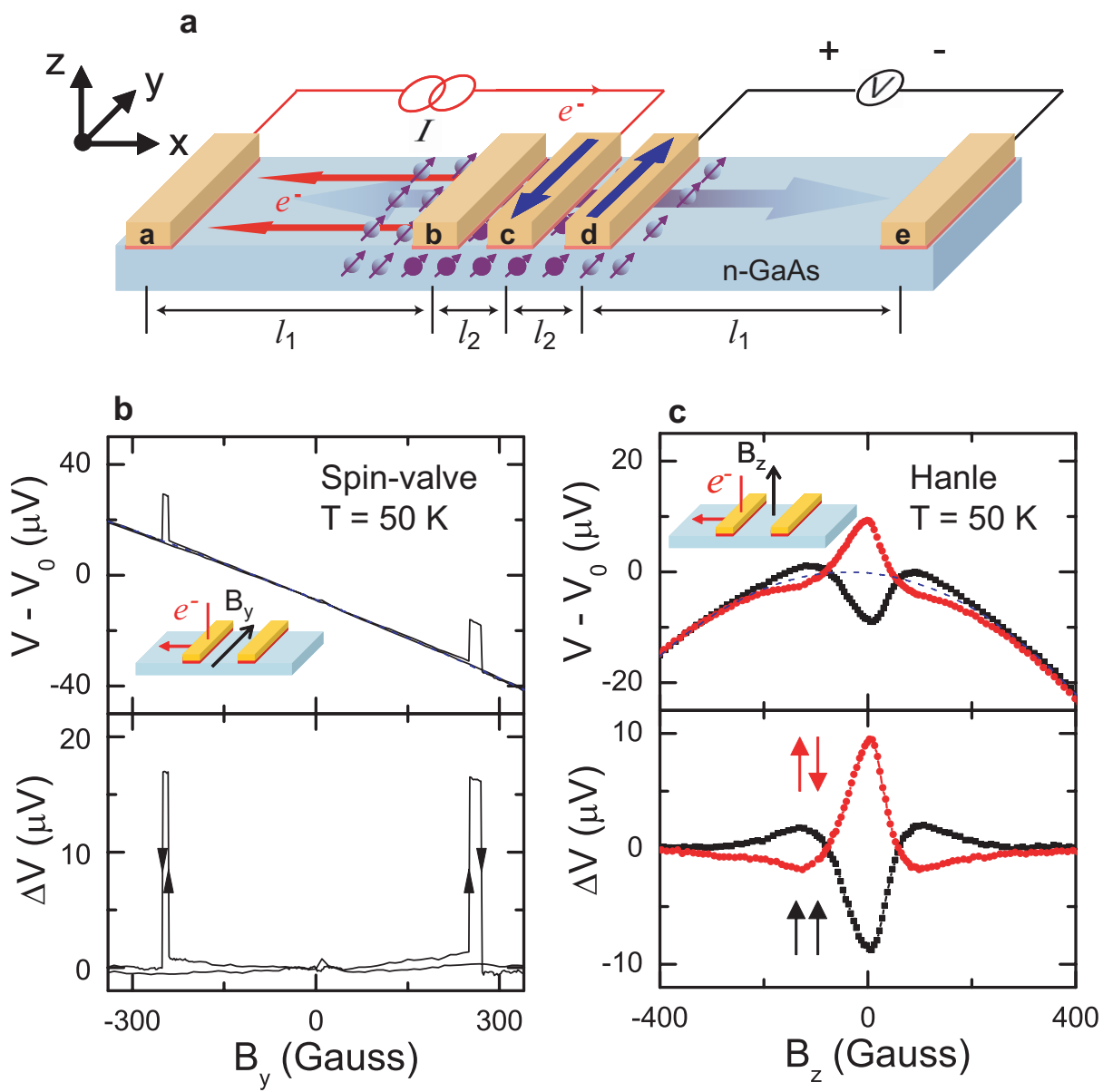

Figure 1, Lou et al.

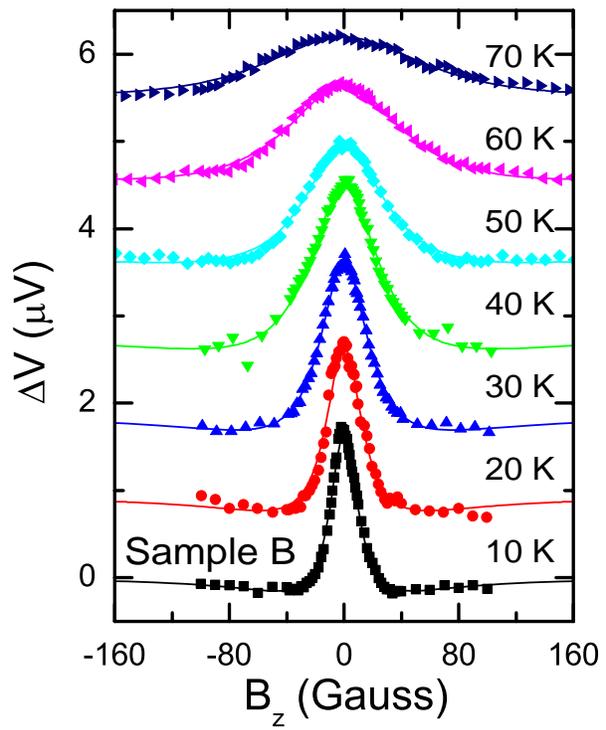

Figure 2, Lou *et al.*

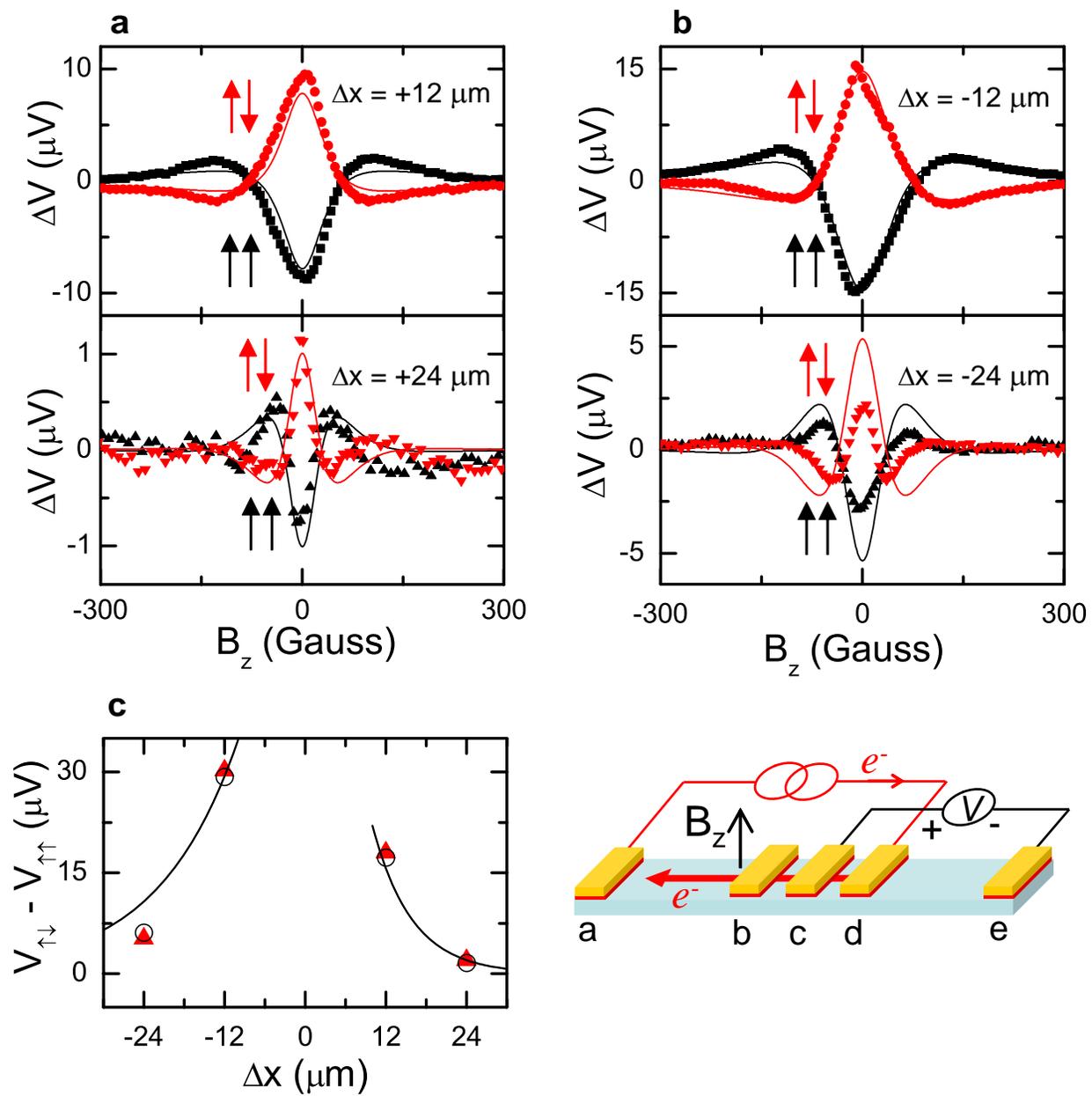

Figure 3, Lou *et al.*

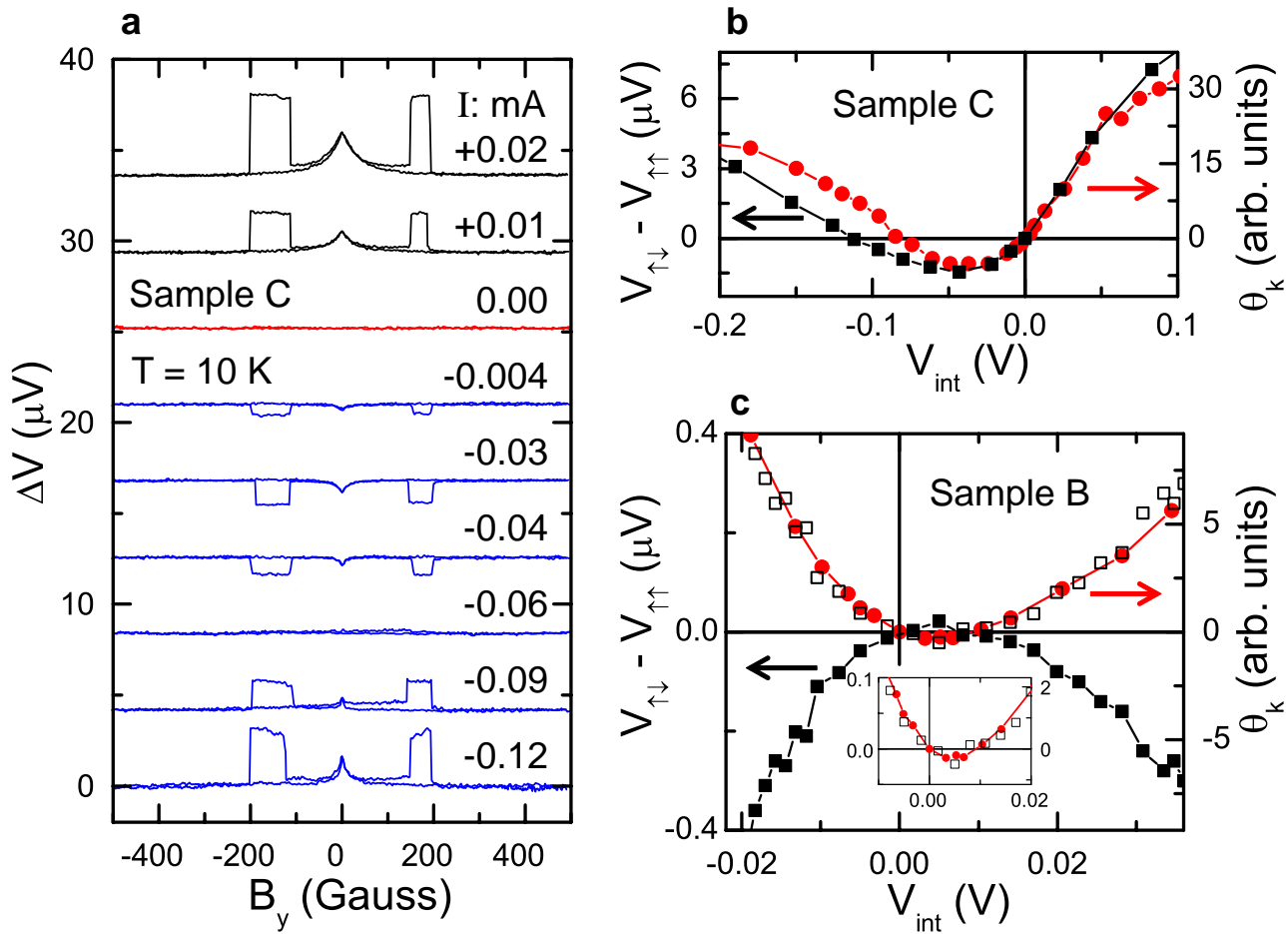

Figure 4, Lou *et al.*

# Lou *et al.*, Supplementary Information

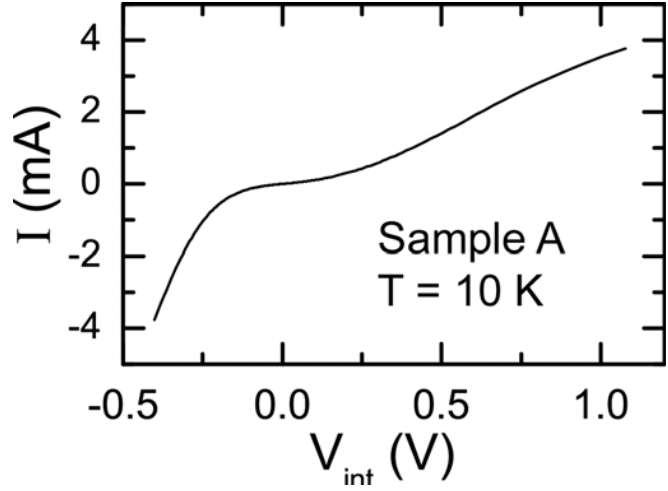

**Supplementary Figure 1**: I-V curve for sample A at 10 K for a current $I_{ab}$ and voltage $V_{cb}$ (See Fig. 1 of main text for contact labels). The voltage $V_{cb}$ corresponds to $V_{int}$, which is the voltage drop between the channel and the source contact. This includes a small contribution from the channel underneath the source contact but is otherwise the best upper bound on the Fe-GaAs interfacial voltage.

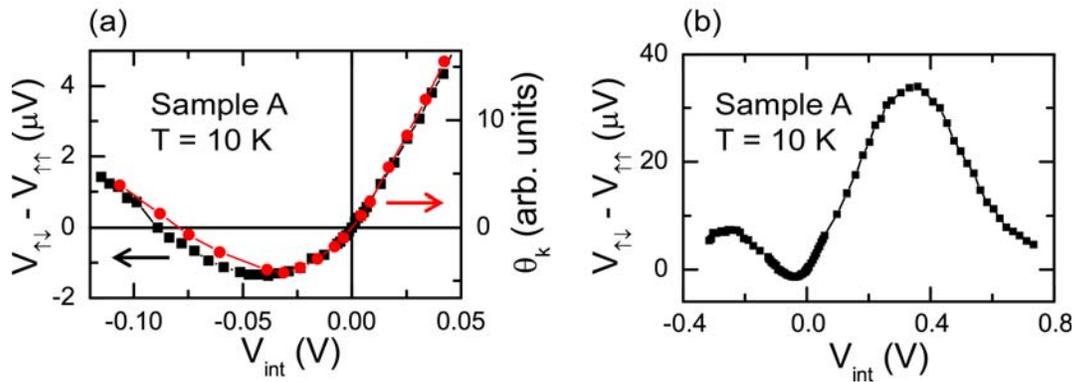

**Supplementary Figure 2:** (a) $V_{\uparrow\downarrow} - V_{\uparrow\uparrow}$ (black squares, left axis) for sample A near zero bias is shown vs. the interfacial voltage $V_{int}$ at the source contact. The Kerr rotation $\theta_K$ (red circles, right axis) vs. $V_{int}$ measured on the same device is shown as red circles. These data correspond to those shown in Fig. 4(b) and (c) in the main text for samples B and C. (b) $V_{\uparrow\downarrow} - V_{\uparrow\uparrow}$ vs $V_{int}$ over the entire bias range of the experiment. As in the main text, positive $V_{int}$ corresponds to injection of electrons from Fe into GaAs. The suppression of $V_{\uparrow\downarrow} - V_{\uparrow\uparrow}$ at large positive $V_{int}$ is accompanied by a decrease in the spin lifetime $\tau_s$ as determined from Hanle curves. The decrease in $\tau_s$ is the primary reason for the suppression of the signal at high bias.